\documentclass[preprint,3p,review,12pt]{elsarticle}

\biboptions{sort&compress}
\usepackage{epsfig}
\graphicspath{{Figure/}}
\usepackage{booktabs}
\usepackage{natbib}
\usepackage{amssymb}
\usepackage{graphicx,epstopdf}
\usepackage[pdf]{pstricks}
\usepackage{subcaption}
\usepackage{amsmath}
\usepackage{float}
\usepackage{xcolor}
\usepackage{tabularx}
\usepackage{verbatim}
\usepackage{lineno}

\makeatletter
 \providecommand{\url}[1]{%
   \begingroup
     \let\bibinfo\@Secondoftwo
     \urlstyle{rm}%
     \href{http://dx.doi.org/#1}{%
       \discretionary{}{}{}%
       \nolinkurl{#1}%
     }%
   \endgroup
 }
 \makeatother

\usepackage{hyperref}
\hypersetup{
    colorlinks=true,
    linkcolor=blue,
    filecolor=blue, 
    citecolor=blue,
    urlcolor=blue,
    pdftitle={Manuscript},
    pdfpagemode=FullScreen,
    }
\usepackage{nameref}  

\begin{document}

\begin{frontmatter}



\title{Non-Monotonic Marangoni Suppression of Hydrodynamic Coarsening in Bicontinuous Liquid-Liquid Phase Separation}


\author[inst1]{Tian Liu}
\author[inst1]{Haohao Hao}
\author[inst1]{Jiaxi Liu}
\author[inst1]{Yongjie Zhou}
\author[inst1]{Feiyu An}
\author[inst1]{Huanshu Tan\corref{cor1}}
\ead{tanhs@sustech.edu.cn}
\cortext[cor1]{Corresponding author}

\date{}

\affiliation[inst1]{organization={Multicomponent Fluids group, Center for Complex Flows and Soft Matter Research \& Department of Mechanics and Aerospace Engineering, Southern University of Science and Technology},
            city={Shenzhen},
            postcode={518055}, 
            state={Guangdong},
            country={China}}

\begin{abstract}

Hydrodynamic coarsening of bicontinuous domains is a central process in liquid-liquid phase separation, yet how soluble surfactants regulate this process remains poorly understood. 
Using a validated two-order-parameter phase-field model coupled to the incompressible Navier--Stokes equations, we show that hydrodynamic coarsening is suppressed primarily by surfactant-induced Marangoni stresses rather than by the reduction of mean interfacial tension alone. 
These stresses hinder interfacial coalescence, reorganize the local vortical flow, and thereby redirect the morphological evolution of bicontinuous domains. 
A central result is that this suppression depends non-monotonically on the surfactant Péclet number, with the strongest inhibition occurring at an intermediate value, $Pe_\psi=10$, rather than at $Pe_\psi=1$ or 100.
Analyses of force evolution, interfacial surfactant statistics, and decomposed surfactant flux budgets show that this non-monotonicity arises from a competition between surfactant replenishment and gradient retention. 
At low $Pe_\psi$, diffusion efficiently replenishes the interface but smooths interfacial concentration gradients; at high $Pe_\psi$, advection preserves interfacial heterogeneity but leaves the interface insufficiently supplied with surfactant. 
The strongest suppression therefore occurs when sufficient interfacial surfactant loading coexists with persistent concentration gradients. 
These results establish a transport-controlled mechanism by which soluble surfactants regulate bicontinuous hydrodynamic coarsening.
 
\end{abstract}



\begin{keyword}
Marangoni Effect \sep Liquid-liquid Phase Separation \sep Hydrodynamics Coarsening 
\end{keyword}

\end{frontmatter}


\section{Introduction}
\label{sec:Introduction}

Liquid-liquid phase separation~(LLPS) is a ubiquitous pathway for generating spatially heterogeneous fluid structures, arising in systems that range from everyday physicochemical processes~\citep{vitale2003liquid,tan2016evaporation} to biological organization~\citep{Hyman2014,wang2021liquid,Alberti2019} and diverse engineering applications~\citep{lohse2020physicochemical, mahboob2022eor,ravera2021emulsification}. 
By tuning thermodynamic control parameters, such as temperature and solution composition, one can systematically regulate both the size and connectivity of the emerging domains~\citep{Cates_Tjhung_2018,ravera2021emulsification}, thereby inducing transitions between droplet-like and bicontinuous morphologies~\citep{laradji1991dynamics,komura1997two,mahboob2022eor,Cates_Tjhung_2018}. 
This tunability has made LLPS a versatile route for fabricating complex micro- and mesoscale structures, including microemulsions~\citep{LI2023299, mahboob2022eor}, porous polymers~\citep{tree2017multi}, well-defined microparticles~\citep{alhasan2023nonsolvent, tan2019porous, DELUCA2024569}, as well as Janus-type~\citep{zhang2022janus,yuan2017phase} and core-shell droplets~\citep{hester2023}.

Over the past decades, substantial progress has been made in understanding domain coarsening during LLPS in multicomponent fluid systems~\citep{Cates_Tjhung_2018,ravera2021emulsification,zwicker2025physics}. 
A central result is the emergence of dynamic scaling laws that relate the characteristic domain size $R$ to the elapsed time $t$ through a power law, $R(t)\sim t^\alpha$~\citep{Bray01061994}. 
In the absence of hydrodynamics, the dynamics are described by Model B, which predicts diffusion-limited coarsening with exponent $\alpha=1/3$.
When hydrodynamic interactions become important, the system crosses over to Model H, in which advective transport governs the late-stage evolution. 
In this regime, the growth exponent approaches $\alpha=1$ in the capillary-viscous regime~\citep{siggia1979late}, crosses over to $\alpha=2/3$ in the capillary-inertial regime~\citep{furukawa1985effect,kendon2000scaling}, and is further reduced to $\alpha\leq 1/2$ in the asymptotic regime~\citep{grant1999spinodal}. 
These classical scaling results, however, rely on a local-equilibrium picture and become less transparent once interfacial transport, flow, and composition remain strongly coupled over evolving diffuse interfaces. 
The addition of surfactants introduces precisely such a coupling. 
Surfactant molecules continuously exchange between the bulk and the interface through diffusion, flow-driven convection, and adsorption-desorption kinetics, leading to a dynamically evolving interfacial tension that couples thermodynamics, interfacial transport, and hydrodynamics~\citep{Engblom2013,LIU20109166,SOLIGO20191292,YANG2022110909,yang2018numerical,manikantan2020surfactant}.

In surfactant-laden multiphase flows, this transport breaks the uniform distribution of surfactants along the interface and generates spatial variations in interfacial concentration. 
These variations produce tangential gradients in interfacial tension and hence Marangoni stresses, which actively modify interfacial mobility and reorganize the surrounding flow field~\citep{ravera2021emulsification,HAO2026114510, manikantan2020surfactant, skartlien2012coalescence, Dai2008, Pan_Tseng_Chen_Huang_Wang_Lai_2016,SOLIGO20191292, HAO2025114058}. 
A representative example is the mutual approach of two surfactant-covered droplets, where lubrication-driven drainage of the intervening thin film redistributes surfactant and creates strong interfacial concentration gradients. 
The resulting Marangoni stresses oppose interfacial motion, retard film drainage, and prolong the lifetime of the thin liquid film~\citep{ravera2021emulsification, manikantan2020surfactant, skartlien2012coalescence,krebs2012coalescence, Dai2008, Pan_Tseng_Chen_Huang_Wang_Lai_2016,SOLIGO20191292}. 
Consistent with this mechanism, extensive experimental and numerical studies have shown that surfactants can retard droplet coalescence~\citep{skartlien2012coalescence,krebs2012coalescence} and stabilize microemulsion structures~\citep{pal2019phase,ravera2021emulsification}. 
Despite these advances, a mechanistic understanding of how Marangoni stresses regulate interfacial dynamics and statistically influence bicontinuous domain coarsening during LLPS remains incomplete. 
In particular, it remains unresolved how soluble-surfactant transport controls Marangoni stresses during hydrodynamic coarsening, and why this control can become non-monotonic as the balance between diffusion and convection changes.

To describe surfactant effects theoretically, early continuum studies introduced free-energy formulations with two coupled order parameters, representing the compositional phase field and the surfactant concentration, respectively, thereby establishing a foundation for analyzing surfactant-mediated phase separation and coarsening~\citep{laradji1991dynamics,komura1997two,nekovee2000lattice, SOLIGO20191292}. 
Building on this framework~\citep{SOLIGO20191292}, we consider a symmetric binary fluid mixture with soluble surfactants and neglect surface viscoelastic effects~\citep{kim2013effect,manikantan2020surfactant}. 
Within this setting, we investigate how surfactant transport regulates hydrodynamic coarsening of bicontinuous domains. 
Our focus is not merely on whether surfactants slow coarsening, but on how transport-controlled Marangoni stresses alter the coarsening pathway and why this suppression depends non-monotonically on the surfactant Péclet number, $Pe_\psi$. 
In particular, we show that the dominant effect arises from Marangoni stresses rather than from the reduction of mean interfacial tension alone, and we identify the transport competition that determines the strength of this Marangoni-mediated suppression.

The remainder of the paper is organized as follows. 
Section~\ref{sec2} presents the phase-field formulation and governing equations for LLPS with soluble surfactants. 
Section~\ref{sec3} validates the numerical implementation against established coarsening scalings and a benchmark problem of surfactant-laden droplet deformation. 
Section~\ref{sec4} examines how surfactant-induced Marangoni stresses modify phase morphology and suppress hydrodynamic coarsening. 
Section~\ref{sec5} analyzes the non-monotonic influence of the surfactant Péclet number and identifies its transport origin. 
Finally, Section~\ref{sec6} summarizes the main findings and discusses their implications for controlling microstructure evolution in complex fluid systems.

\section{Phase-field formulation and numerical methodology}
\label{sec2}
\subsection{Free-energy model and equilibrium surfactant partitioning}

To model LLPS, we adopt a two-order-parameter time-dependent Ginzburg-Landau free-energy functional, $F(\phi, \psi)= \int_{\Omega} f_\phi(\phi)+f_\psi(\psi)+f_a(\phi,\psi)+f_{ex}(\phi,\psi) d \Omega$ (\ref{sec:sample:appendixa}), which has previously been employed in phase-field studies of surfactant-laden droplets coalescence~\citep{SOLIGO20191292}.
In the present LLPS simulations, interfaces emerge spontaneously from an initially homogeneous mixture and subsequently evolve under the coupled effects of thermodynamics, hydrodynamics, and surfactant transport.
Although the governing equations share the same formal structure, the physical setting and modeling objectives differ; the formulation is therefore briefly reintroduced here for clarity and completeness.

The phase field $\phi(\mathbf{x},t)$ and the soluble surfactant concentration $\psi(\mathbf{x},t)$ are governed by two convective Cahn-Hilliard equations, respectively, which in dimensionless form can be written as follows~\citep{SOLIGO20191292,Engblom2013,LIU20109166,YANG2022110909,van2006diffuse}
\begin{equation}\label{222}
  \frac{\partial \phi}{\partial t} + \mathbf{u} \cdot \nabla \phi=\frac{1}{Pe_\phi} \nabla \cdot ( M(\phi) \nabla \mu_\phi)+\zeta,
\end{equation}
\begin{equation}\label{333}
  \frac{\partial \psi}{\partial t} + \mathbf{u} \cdot \nabla \psi=\frac{1}{Pe_\psi} \nabla \cdot (M(\psi) \nabla \mu_\psi),
\end{equation}
where \(\mathbf{u}\) is the fluid velocity, $M(\phi)$ and $M(\psi)$ denote mobilities, and \textcolor{black}{$\zeta$ is an initial random noise field introduced to trigger phase separation.}
Unless otherwise specified, we take $M(\phi) = 1$ and $M(\psi)=\psi (1-\psi)$~\citep{Engblom2013}.
The P\(\acute{\text{e}}\)clet numbers $Pe_\phi$ and $Pe_\psi$ quantify the ratio of convective to diffusive transport of $\phi$ and $\psi$, respectively.
The corresponding chemical potentials $\mu_\phi$ and $\mu_\psi$ are obtained as variational derivatives of the free energy functional $F(\phi, \psi)$ with respect to $\phi$ and $\psi$, respectively
\begin{equation}\label{444}
  \mu_\phi=\frac{\delta F}{\delta\phi}=\phi(\phi^2-1)-Cn^2\Delta\phi + \underbrace{4S_{ad}\psi\phi(1-\phi^2)+2E_x\psi\phi}_{\mu_\phi^{\prime}},
\end{equation}
\begin{equation}\label{555}
  \mu_\psi =\frac{\delta F}{\delta \psi} = Pi\ln(\frac{\psi}{1-\psi})-S_{ad}(1-\phi^2)^2 +E_x \phi^2.
\end{equation}
The term $\mu_\phi^{\prime}$ represents the surfactant contribution to the chemical potential~$\mu_\phi$, and thus describes how the surfactant concentration distribution influences the diffusion of $\phi$ during phase separation.
We neglect this coupling by setting $\mu_\phi^{\prime}=0$, implying that the diffusion of $\phi$ is assumed to be independent of the surfactant concentration.
The temperature-dependent coefficient $Pi$ modulates surfactant diffusivity, while the positive coefficients $S_{ad}$ and $E_x$ are determined by the Langmuir adsorption isotherm (refer to \ref{sec:sample:appendixa}).



At equilibrium, the chemical potentials $\mu_\phi$ and $\mu_\psi$ are constant throughout the entire domain~\citep{Engblom2013,LIU20109166,SOLIGO20191292}. 
Solving Eqs.~\eqref{444} and \eqref{555} under this condition yields the equilibrium profiles
\begin{equation}\label{15}
  \begin{aligned}
   &\phi^{e}(\mathbf{x})=\tanh \left(\frac{\mathbf{x}}{\sqrt{2}Cn} \right),
   &\psi^{e}(\mathbf{x})=\frac{\psi_b}{\psi_b+\psi_c(\phi)(1-\psi_b)}, 
 \end{aligned}
\end{equation}
where $\psi_c(\phi)$ is defined as $\psi_c(\phi)=\exp\left(-\frac{1-\phi^2}{Pi}\left[S_{\mathrm{ad}}(1-\phi^2)+E_x\right]\right)$,  and represents an effective equilibrium partition function that encodes the adsorption affinity of surfactants to the interface. 
Specifically, $\psi_c(\phi)$ acts as a $\phi$-dependent factor, with the term $(1-\phi^2)$ localizing the adsorption effect near the diffuse interface ($\phi \approx 0$), while $S_{\mathrm{ad}}$ and $E_x$ respectively characterize the adsorption strength and intermolecular interaction energy. 
As a result, $\psi_c(\phi)$ governs the equilibrium redistribution of surfactants between the bulk and the interface.

Under the dilute-surfactant assumption ($\psi_b \ll 1$), the equilibrium interfacial surfactant concentration (at $\phi = 0$) reduces to Langmuir adsorption isotherm
\begin{equation}\label{14}
  \psi^{e}_{\mathrm{i}} = \frac{\psi_b}{\psi_b+\psi_{ci}},
\end{equation}
where $\psi_{ci} = \exp\left(-(S_{ad}+E_x)/Pi\right)$ denotes the effective adsorption parameter evaluated at the interface.
In the dilute limit ($\psi_b \ll \psi_{ci}$), this expression further simplifies to
\begin{equation}
  \psi^{e}_{\mathrm{i}} \approx \frac{\psi_b}{\psi_{ci}}.
\end{equation}
This relation shows that the interfacial surfactant concentration scales linearly with the bulk concentration, with $\psi_{ci}^{-1}$ acting as an effective partition coefficient. 
Physically, larger values of $S_{ad}$ or $E_x$ decrease $\psi_{ci}$, thereby increasing $\psi^{e}_{\mathrm{i}}$ and promoting stronger accumulation of surfactants at the interface.

\subsection{Hydrodynamics coupling and Marangoni forcing}

We consider a symmetric binary fluid mixture in which the two components have identical density and viscosity, i.e, $\rho=\rho_1=\rho_2$ and $\eta=\eta_1=\eta_2$.
To capture the hydrodynamic coarsening dynamics during LLPS, we further couple the Navier-Stokes equation to the phase-field formulation, and its dimensionless form is given as follows~\citep{SOLIGO20191292}
\begin{equation}\label{666}
  \begin{aligned}
    \nabla \cdot \mathbf{u} =0,
  \end{aligned}
\end{equation}
\begin{equation}\label{777}
  \begin{aligned}
     \frac{\partial \mathbf{u}}{\partial t}+\mathbf{u} \cdot \nabla \mathbf{u} &=-\nabla p+\frac{1}{Re} \Bigg[ \nabla^2 \mathbf{u} 
    +\underbrace{\frac{1}{C a}\frac{3C n}{\sqrt{8}} \Big( \nabla \cdot \overline{\boldsymbol{\tau}}+ \beta_s \ln (1-\psi) \nabla \cdot \overline{\boldsymbol{\tau}}\Big)}_{\text {Capillary force},~\mathbf{F_{Ca}}}\\
    & \quad +\underbrace{Ma\frac{3Cn}{\sqrt{8}}  \nabla \ln (1-\psi) \cdot \overline{\boldsymbol{\tau}}}_{\text {Marangoni force},~\mathbf{F_{Ma}}}\Bigg]
  \end{aligned}
\end{equation}
where $p$ is the pressure and $\overline{\boldsymbol{\tau}}=|\nabla \phi|^2 \mathbf{I}-\nabla \phi \otimes \nabla \phi$ is the Korteweg stress tensor~\citep{korteweg1901forme}. 
The other dimensionless parameters involves the Reynolds number~$Re = \rho U L / \mu$, the capillary number~$Ca = \mu U / \sigma_0$, and the Marangoni number~$Ma=\beta_s/Ca = \beta_s \sigma_0 / \mu U$, where $U$, $L$, and $\sigma_0$ denote the characteristic velocity, length, and surface tension coefficient, respectively.
The choices of these hydrodynamic parameters are summarized in~\ref{sec:sample:appendixb}.
The last two terms on the right side of Eq.~\eqref{777} correspond to the normal and tangential components of the interfacial tension force—namely, the capillary force and the Marangoni force.
To account for the influence of surfactants on interfacial tension, we adopt the Langmuir equation of state, $f_\sigma(\psi)=1+\beta_s \ln(1-\psi)$, where $\beta_s$ the elasticity number that quantifies the sensitivity of the interfacial tension to variations in local surfactant concentration.
\textcolor{black}{In practice, a lower bound $f_\sigma(\psi) \geqslant 0.5$ is imposed to prevent unphysical surface-tension reductions predicted by the equation of state at high surfactant concentrations~\citep{SOLIGO20191292,manikantan2020surfactant}.}

Depending on whether hydrodynamics and soluble surfactants are included, we consider three numerical modes of increasing physical complexity.
(1)~Model B is a purely diffusive model in which only the compositional order parameter  $\phi$ evolves according to the Cahn-Hilliard equation, Eq.~\eqref{222}.
(2)~Model H extends Model B by incorporating hydrodynamics through the Navier-Stokes equations, thereby accounting for hydrodynamic coarsening effects~\citep{hohenberg1977theory,Cates_Tjhung_2018}.
(3)~Model HS further extends Model H by including soluble surfactants and their coupling to compositional and flow fields, enabling investigation of surfactant-regulated  hydrodynamic coarsening.
In Model HS, surfactant effects arise from both capillary forces associated with interfacial tension variations and Marangoni forces induced by interfacial-tension gradients (Eq.~\eqref{777}).
In the absence of surfactants (Models B and H), the free energy reduced to $F(\phi) = \int_{\Omega}{f_\phi(\phi)} ~ d{\Omega}$, consistent with the classical description of phase separation~\citep{hohenberg1977theory,Cates_Tjhung_2018}.


\subsection{Numerical implementation and setup}

We employ a finite-volume method on a rectangular uniform marker-and-cell (MAC) mesh to spatially discretize the two convective Cahn-Hilliard equations (Eqs.\eqref{222} and \eqref{333}) and the momentum equation Eq.\eqref{777}. 
In this framework, vector quantities (i.e., the velocity $\mathbf{u}$ and the surface tension force) are stored at cell faces, whereas scalar quantities (i.e., $\phi$, $\psi$, $\mu_\phi$, $\mu_\psi$, $p$, $\rho$, and $\eta$) are defined at cell centers.
The advection terms in the two convective Cahn-Hilliard equations are discretized using a fifth-order weighted essentially nonoscillatory (WENO) scheme in spatial, with the local flow velocity determining the up-winding direction~\citep{Shu1998}.
Unlike Ref.~\citep{SOLIGO20191292}, where IMplicit-EXplicit (IMEX) scheme are adopted, these two equations in the present work are advanced explicitly in time for simplicity.
Specifically, the advection terms are advanced using a second-order Adams-Bashforth scheme, while the diffusion terms are treated with a first-order forward Euler scheme.
For the momentum equation Eq.\eqref{777}, the advection term is advanced using the second-order Adams-Bashforth method, while the viscous term is treated with a Cran-Nicolson scheme.
The standard projection method is used to solve the coupled equations Eq.\eqref{666} and Eq.\eqref{777}, where the Poisson equation is iteratively solved using the Gauss-Siedel method. 
Further numerical details can be found in Refs.~\cite{ding2007diffuse, HAO2024104750}.

Here, the initial conditions for two order parameters are prescribed as follows: for the phase field, $\xi(\mathbf{x}) = \phi_0 + 0.001 \mathrm{rand}(\mathbf{x})$, where the initial volume fraction $\phi_0 =  0$ and the random perturbation satisfies $\sum_{\Omega} \mathrm{rand}(\mathbf{x}) = 0$; 
and for the surfactant concentration, $\psi_0(\mathbf{x}) = \psi_b$.
Unless otherwise specified, the computational domain is $[0,2\pi] \times [0,2\pi]$ with a uniform grid spacing of $dx = dy = 2\pi/300$, and periodic boundary conditions are applied to all variables on all boundaries.


\section{Validation against coarsening scalings and surfactant-laden droplet deformation}
\label{sec3}
In this section, we validate the accuracy of the phase-field model and its numerical implementation. 
We first compare the simulated growth rate of the domain size $R(t)$ during the coarsening of bicontinuous structures with classical scaling laws~\citep{Bray01061994,siggia1979late,furukawa1985effect,kendon2000scaling,grant1999spinodal}. 
Then, we simulate the deformation of a surfactant-laden droplet in a shear flow.
\subsection{Domain growth in Model B and Model H}
\label{sec3.1}
We perform numerical simulations using both Model B and Model H. Model B, which solves the Cahn-Hilliard equation without hydrodynamics, serves as a benchmark for purely diffusive dynamics. In contrast, Model H incorporates the coupled Navier-Stokes equations to account for hydrodynamic effects. 
The parameters for each model are as follows: for Model B, the mobility is $M_\phi=10^{-3}$ with a time step of $dt = 1 \times 10^{-4}$; for Model H, the Reynolds and capillary numbers are set to $Re=0.01$ and $Ca=1$, respectively, with a smaller time step of $dt = 1 \times 10^{-5}$.
\begin{figure}[H]
  \centering
  \includegraphics[width=1.0\textwidth]{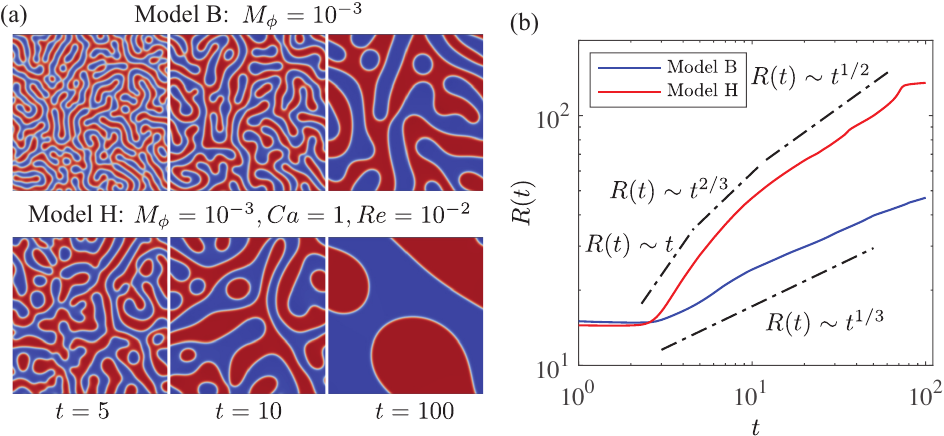}
  \caption{
  (a) Phase patterns from Model B (top row) and Model H (bottom row) at times~$t=5, 10$ and 100.
  (b) Temporal evolution of the domain size $R(t)$ for both models at a mobility of $M_\phi=10^{-3}$ and Reynolds number of $R(t)=0.01$.
}
  \label{fig:1}
\end{figure}

Figure~\ref{fig:1}(a) compares the coarsening behavior during liquid-liquid phase separation (LLPS) between the purely diffusive regime (Model B, top row) and the flow-driven regime (Model H, bottom row). 
The phase patterns at $t=5,10$ and 100 clearly illustrate that hydrodynamic effects significantly accelerate the domain coarsening process, leading to much larger phase structures in Model H at equivalent times.

To quantitatively evaluate the simulations against theory, Figure~\ref{fig:1}(b) presents the temporal evolution of the domain size $R(t)$, 
\textcolor{black}{obtained from the inverse first moment of the circularly averaged structure factor~$S(k, t)$, as~\cite{nekovee2000lattice,kendon2000scaling}
\begin{equation}
R(t)=\frac{2 \pi \int S(k, t) d k}{\int  k S(k, t) d k},
\end{equation}
where $k$ denotes the wave number.}
The results show that the growth rate of $R(t)$ in Model B agrees well with the Lifshitz-Slyozov-Wagner (LSW) theory~\citep{LIFSHITZ196135,wagner1961theory} for diffusion-dominated coarsening, following the scaling law $R(t) \sim t^{1/3}$.
\textcolor{black}{In contrast, Model~H simulations clearly resolve three successive coarsening regimes: a viscous hydrodynamic regime with~$R(t)\sim~t$~\cite{siggia1979late,bray2002theory} for $t<5$, an inertial hydrodynamic regime with~$R(t)\sim t^{2/3}$~\citep{furukawa1985effect,kendon2000scaling} for $5\leqslant t<10$, and a asymptotic regime characterized by a reduced growth exponent~$R(t)\sim~t^{1/2}$~\citep{grant1999spinodal}for $ t\geqslant 10$.
The good agreement between these numerical results and established theoretical scaling laws validates the accuracy of the present model for simulating LLPS dynamics across diffusive and hydrodynamic regimes.}

\subsection{Surfactant-laden droplet in a shear flow}
\label{sec3.2}
\begin{figure}[H]
  \centering
  \includegraphics[width=1.0\textwidth]{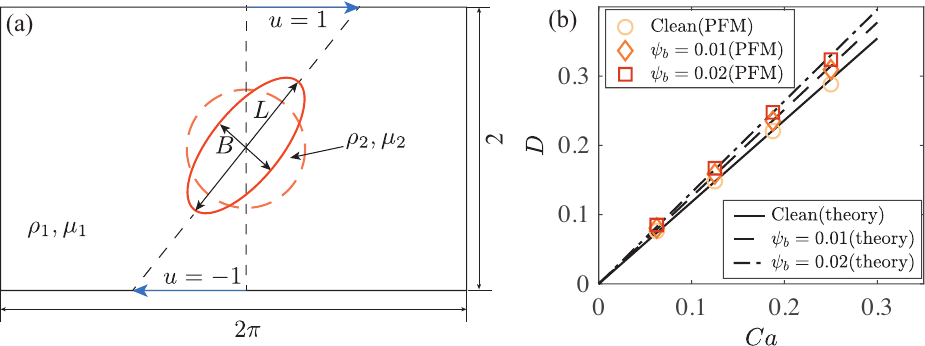}
  \caption{
  (a) Schematic of the computational setup for a surfactant-laden droplet under shear flow. 
  (b) Comparison of Taylor deformation parameter $D$ with analytical results (lines) at different $Ca$ numbers and bulk surfactant concentrations $\psi_b$.}
  \label{fig:2}
\end{figure}

To validate the program for hydrodynamics, we simulate the deformation of a surfactant-laden droplet in shear flow, a benchmark test used in previous work~\citep{SOLIGO20191292}.
As illustrated in Fig.~\ref{fig:2}(a), a droplet with diameter $d=0.8$ is placed at the center of the computational domain, which has a size of $2\pi \times 2$.
The shear flow is driven by imposing velocities $u = 1$ and $u = -1$ on the top and bottom boundaries (blue arrows in Fig.~\ref{fig:2}(a)), respectively, while  periodic conditions are applied on the left and right sides.
The initial distributions of the phase field $\phi$ and surfactant concentration $\psi$ are set in equilibrium using Eq.\eqref{15}. 
Surfactant profile parameters in Eq.\eqref{555} are chosen as $Pi=1.25$, $S_{ad}=0.5$, and $E_x=4.2735$, consistent with previous work~\citep{SOLIGO20191292}.
Other dimensionless parameters are set as follows: $Pe_\phi=150$, $Pe_\psi=100$, $Re=0.1$, $\beta_s=0.5$, $dx=0.01$, and $dt=1 \times 10^{-4}$.

The drop deformation is quantified by the Taylor deformation parameter $D=(L-B)/(L+B)$, where $L$ and $B$ denote the major and minor axes, respectively, as shown in Fig.~\ref{fig:2}(a). 
In the limit of small capillary and Reynolds numbers~($Ca<<1$ and $Re<<1$), the Taylor deformation parameter $D$ is governed by $Ca$ and can be described by an analytic relation that accounts for wall effects~\citep{SHAPIRA1990305}, as
\begin{equation}\label{s3.2}
  \begin{aligned}
 & D=\frac{35}{32}Ca\left[ 1+C_{SH}\frac{3.5}{2}\left(\frac{d}{4h}\right)^{3}\right], \\
   \end{aligned}
\end{equation}
where $C_{SH} = 5.6996$~\citep{SHAPIRA1990305} and $h$ is half-width of the computational domain.
To compare the simulation results against this analytical relation, the theoretical deformation is computed using an effective capillary number~\citep{SOLIGO20191292}, $Ca_e =(\sigma_0/\sigma_{av})Ca$, where $\sigma_{av}$ is the average surface tension that accounts for the reduction in surface tension due to the presence of surfactants~\citep{Stone_Leal_1990}.

Figure~\ref{fig:2}(b) shows the comparison of Taylor deformation parameter $D$ with analytical results (lines) at different $Ca$ numbers and bulk surfactant concentrations $\psi_b$.
The results show good agreement at low capillary numbers ($Ca<0.2$), while a clear deviation from the theoretical solution emerges as $Ca$ increases.
This is because an assumption of $Ca<<1$ for this theory can be not satisfied.

\section{Marangoni suppression of bicontinuous hydrodynamic coarsening}
\label{sec4}

\begin{figure}
  \centering
  \includegraphics[width=1.0\textwidth]{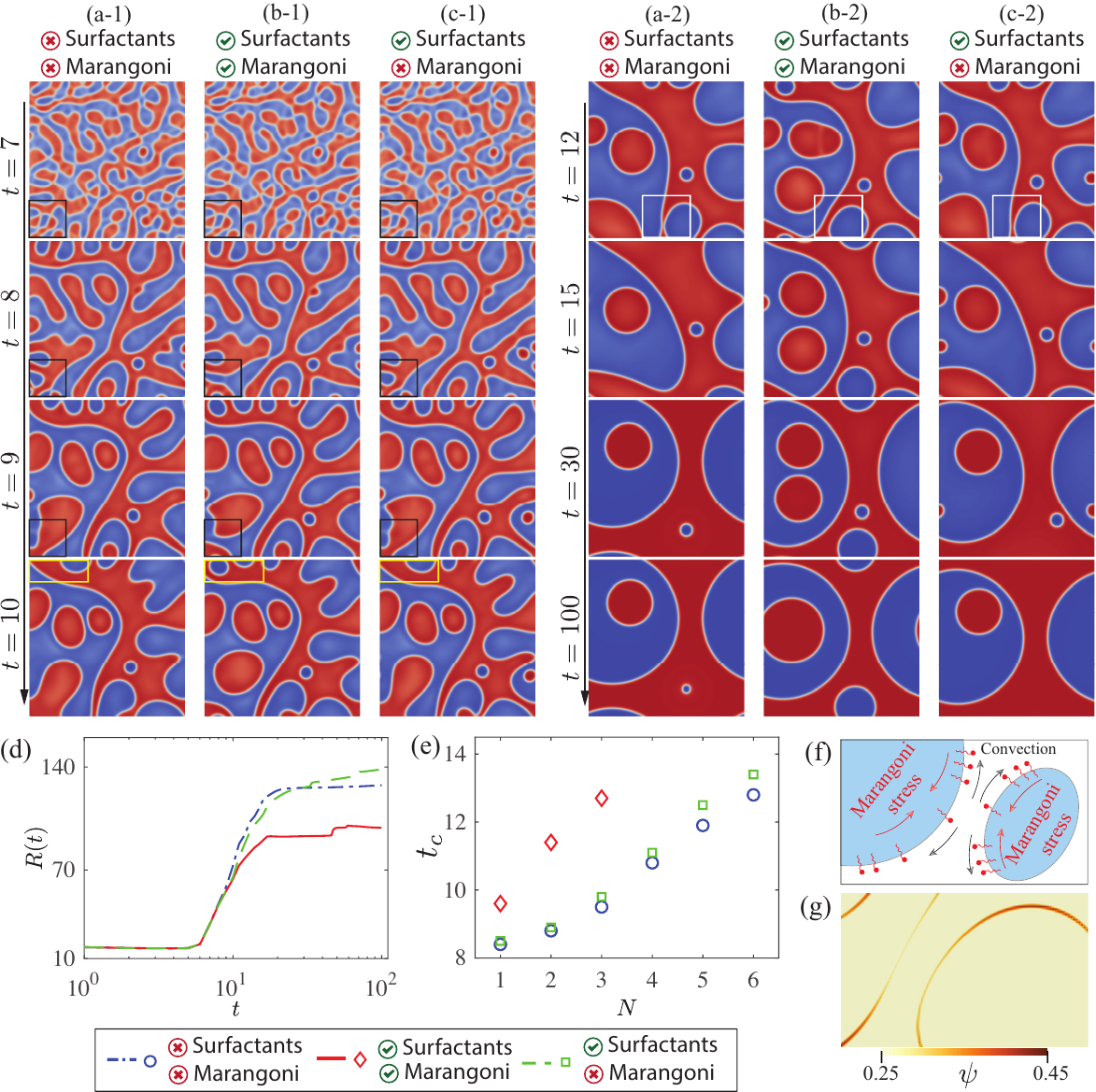}
  \caption{Pattern evolution and Marangoni effects during hydrodynamic coarsening at $Pe_{\psi=10}$ and $Ma = 9$.
  (a-c) Pattern evolution under three conditions: (a) clean interface (no surfactant, no Marangoni forces), (b) surfactant-laden interface with Marangoni forces, and (c) surfactant-laden interface without Marangoni forces.
  (d) Temporal growth of characteristic domain size~$R(t)$ for the three cases. 
  (e) Number of coalescence events $N$ and its corresponding coalescence time $t_c$  over the interval~$t\in [8, 15]$ for each case.
  (f) Schematic illustration of how Marangoni forces inhibits the drainage of the thin liquid film between approaching interfaces. 
  (g) Surfactant concentration field at the region marked by the white rectangular box in (b) at $t=12$.}
 \label{fgr:3}
\end{figure}

We start by analyzing how surfactants modify the hydrodynamic coarsening behavior of bicontinuous structures during phase separation.
Surfactants adsorbed at the interface not only reduce the local surface tension but also generate Marangoni forces due to surface tension gradients along the interface~\citep{scriven1960marangoni, manikantan2020surfactant}. 
To isolate and quantify these effects, we examine the temporal evolution of the phase morphology and the characteristic domain size $R(t)$ for three representative cases: (i) a clean interface without surfactant (Fig.~\ref{fgr:3}(a)), for which the capillary force term reduces  to $\frac{3Cn}{\sqrt{8}Ca}\nabla \cdot \overline{\boldsymbol{\tau}}$ and Marangoni force term is omitted from Eq.~\eqref{777}, (ii) a surfactant-laden interface with Marangoni force (Fig.~\ref{fgr:3}(b)), and (iii) a surfactant-laden interface in which Marangoni force are deactivated (Fig.~\ref{fgr:3}(c)), corresponding to the removal of the Marangoni force term from Eq.~\eqref{777}.

Here, the initial bulk surfactant concentration is set to $\psi_b=0.050$, and the adsorption constant is $\psi_{ci}=0.035$ ($Pi=0.1$, $S_{ad}=0.3$, $E_x=0.03524$), giving an equilibrium interfacial surfactant concentration of~$\psi^e_i\approx 0.588$ according to (Eq.~\eqref{14}). 
The remaining parameters are set to $Re=0.1$, $Ca = 0.1$, and $Pe_\phi=10^2$. 
Additionally, the identical random noise field $\zeta$ with initial phase field~$\phi_0=0$ at~$t=0$ is used for all simulations to eliminate the influence of initial conditions and allow a direct comparison of their subsequent phase pattern evolution.

Figures~\ref{fgr:3}(a-c) present the numerical snapshots of phase patterns at 8 different instants ($t=7, 8, 9, 10, 12, 15, 30, 100$) for the three representatives.
A direct comparison between Fig.~\ref{fgr:3}(a) and Fig.~\ref{fgr:3}(c) shows that when surfactants are present but Marangoni force are suppressed, the coarsening dynamics remain the same as in the clean case.
Specifically, the pattern evolution in Fig.~\ref{fgr:3}(c) closely resembles that in Fig.~\ref{fgr:3}(a), and the corresponding $R(t)$ curve (green dashed line in Fig.~\ref{fgr:3}(d)) deviates only slightly from the clean-interface result (blue dash-dotted line).
This indicates that, under the present parameter setting, the reduction in capillary force induced by the presence of surfactants -- represented by the term $\frac{3Cn}{\sqrt{8}Ca}\nabla \cdot \overline{\boldsymbol{\tau}}$ in Eq.~\eqref{777} -- plays a relatively minor role in modifying hydrodynamic coarsening compared to the Marangoni effects.
In contrast, when Marangoni force are included, the coarsening behavior changes markedly. 
The growth rate of $R(t)$ significantly reduced when $t \geqslant 9$ (red solid line in Fig.~\ref{fgr:3}(d)), by a noticeable alteration of the phase morphology (Fig.~\ref{fgr:3}(b)).
A similar phenomenon has been observed for asymmetric blends (i.e., $\phi_0 \neq 0$)~\citep{love2001three,skartlien2012coalescence}, although the adsorption isotherms was not considered.
These results demonstrate that the Marangoni force are responsible for suppressing hydrodynamic coarsening.

As shown in Fig.~\ref{fgr:3}(d), the influence of Marangoni force becomes pronounced during the interval~$t\in [9, 15]$.
Within this period, we observe that the presence of Marangoni force suppresses the coalescence of approaching interfaces, as highlighted within the yellow and white rectangular boxes of Fig.~\ref{fgr:3}(a-c) at $t=10$ and $t=12$, respectively.
To quantify this suppression, Figure~\ref{fgr:3}(e) shows the number of coalescence events $N$ as a function of their corresponding coalescence times $t_c$ measured over a time interval $\Delta t$, for the three representatives.
The case with Marangoni force exhibits both fewer coalescence events and delayed coalescence times, confirming that Marangoni force hinder interface merging and thereby suppress hydrodynamic coarsening.

The physical mechanism underlying this inhibition and delay is illustrated in Fig.~\ref{fgr:3}(f). 
As two interfaces approach each other, surfactant is convected from the center of the thinning liquid film to its edge~(black arrows), generating interfacial concentration gradients~\citep{SOLIGO20191292,manikantan2020surfactant}. 
These gradients induce Marangoni force (red arrows) that counteracts the convective flow, thereby suppressing the drainage of the thin liquid film between adjacent interfaces and ultimately delaying their coalescence~\citep{Dai2008,Pan_Tseng_Chen_Huang_Wang_Lai_2016, SOLIGO20191292}. 
This mechanism is further supported by the surfactant concentration field at~$t=12$, extracted from the white box region indicated in Fig.~\ref{fgr:3}(b) and shown in Fig.~\ref{fgr:3}(g), which reveals the emergence of interfacial surfactant gradients responsible for the Marangoni force. 

Besides Marangoni force hinders the coalescence of approaching interfaces~\citep{Dai2008,skartlien2012coalescence, SOLIGO20191292}, we observe that it also alters the local phase-pattern morphology during the early stage of coarsening.
As highlighted by the black boxes in Fig.~\ref{fgr:3}(a-c) for $t=7$, 8, and 9, the presence of Marangoni force leads to differences in the small-scale structures among the three cases.
Although these differences appear minor at early times (e.g., $t=8$), they subsequently amplify and give rise to distinct large-scale morphologies at later stages, as evidenced by the phase pattern at~$t=12$, $t=15$, and $t=30$.

\begin{figure}[H]
  \centering
  \includegraphics[width=1.0\textwidth]{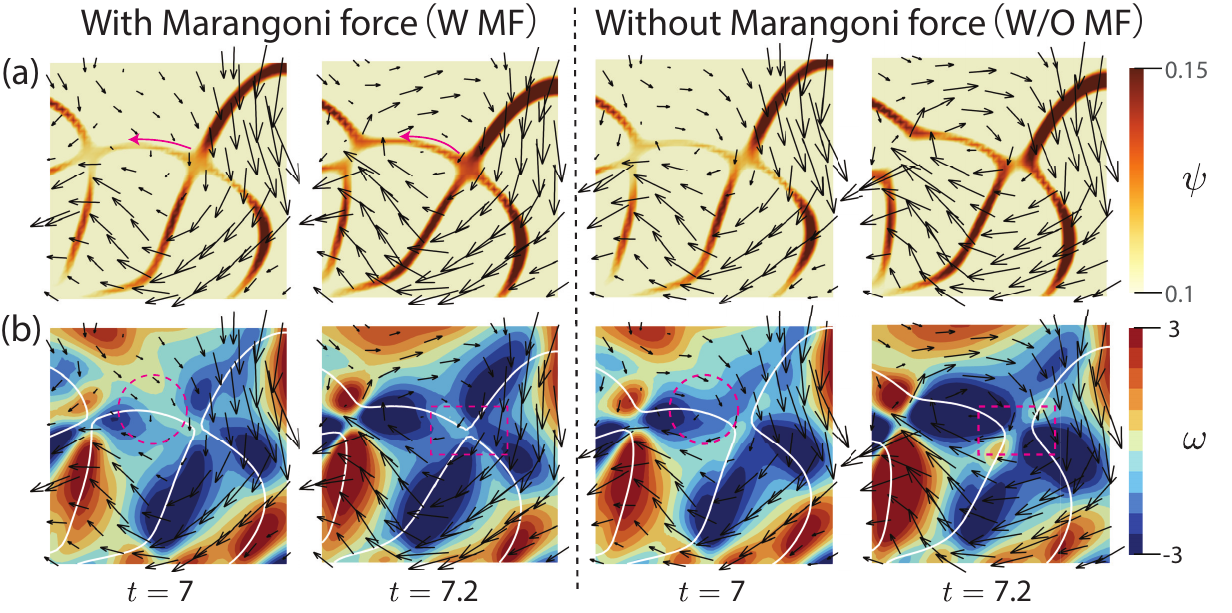}
  \caption{
  Contours of (a) the surfactant concentration and (b) vorticity field with and without Marangoni force at $t=7$ and $t=7.2$.
  The pink arrows in the first two subplots of (a) denote the direction of Marangoni force.
  The black arrows represent the velocity vectors, while the white line in (b) indicates the interface~($\phi = 0$) location.}
  \label{fgr:4}
\end{figure}

To clarify the origin of the early-stage morphological difference, Figure~\ref{fgr:4} compares the surfactant concentration and vorticity fields for the cases with and without Marangoni force at $t=7$ and $t=7.2$.
In there presence of Marangoni force (the first two columns), a non-uniform surfactant distribution develops along the interface at $t=7$, generating tangential Marangoni stresses (pink arrows in Fig.~\ref{fgr:4}(a)).
These stresses do not merely retard interface motion, and they also reorganize the local vortical flow and thereby redirect the morphological pathway of bicontinuous evolution.
As seen from the vorticity contours in Fig.~\ref{fgr:4}(b), the vortex structure highlighted by the pink dashed circle become less coherent in the Marangoni case, and the associated interfacial deformation no longer evolves as a single continuous protrusion.
Instead, by $t=7.2$, the structure inside the pink box splits into two segments.
By contrast, in the absence of Marangoni forces (the last two columns), the corresponding vortex remains more coherent and the surrounding flow retains a more organized rotational pattern.
As a result, the interface continues to deform a continuous manner, and the microstructure develops as an elongated connected feature rather than undergoing splitting, as shown in the fourth column of Fig.~\ref{fgr:4}(b).

These results suggest that Marangoni stresses act as a local flow-selection mechanism, switching the interfacial evolution pathway from coherent extension to fragmentation.
This mechanism is consistent with previous studies showing that Marangoni stresses can regulate vorticity dynamics and interfacial deformation~\citep{Kamat_2020,constante2021role}.

\section{Transport origin of the non-monotonic $Pe_\psi$ dependence}
\label{sec5}

Marangoni stresses are not intrinsic material parameters, but dynamic consequences of surfactant transport. 
We therefore examine how the competition between convective and diffusive transport of surfactants, characterized by the surfactant Péclet number $Pe_\psi$, regulates hydrodynamic coarsening. 
All other parameters are kept identical to those in Sec.~\ref{sec4}.
%

\begin{figure}[h!]
  \centering
  \includegraphics[width=1.0\textwidth]{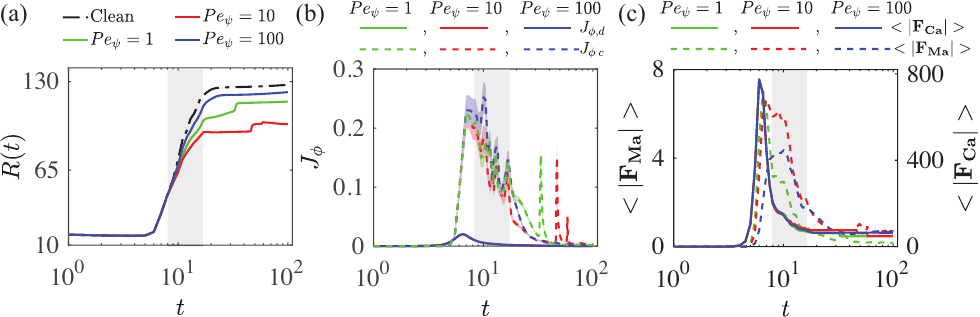}
  \caption{
    Non-monotonic influence of the surfactant P\(\acute{\text{e}}\)clet number $Pe_\psi$ on hydrodynamic coarsening ($Re=~0.1$, $Ca=0.1$, $Pe_\phi=100$).
    (a) Temporal growth of characteristic domain size~$R(t)$ for $Pe_\psi = 1$, 10, and 100.
    The light gray region marks the time interval during which Marangoni forces suppress hydrodynamic coarsening.
	  (b) Temporal evolution of the average convective flux $J_{\phi,c}$ and the average diffusive flux $J_{\phi,d}$ of phase~$\phi$.
	  The light-shaded band represents standard deviation of $J_{\phi,c}$ and $J_{\phi,d}$, and these values are multiplied by 0.15 for clarity.
	  (c) Temporal evolution of the average Marangoni force $<|\mathbf{F_{Ma}}|>$ and the average capillary force $<|\mathbf{F_{Ca}}|>$.
  }
  \label{fgr:5}
\end{figure}

Fig.~\ref{fgr:5}(a) shows the temporal evolution of the characteristic domain size $R(t)$ for $Pe_\psi = 1$ (green solid line), 10 (red solid line), and 100 (blue solid line), together with the clean-interface reference (black dashed line).
In the presence of surfactants, the growth of $R(t)$ is systematically suppressed over the interval $8\leq t \leq 15$ (gray shaded region), indicating a clear deviation from classical hydrodynamic coarsening.
A key observation is that this suppression is non-monotonic in the surfactant P\(\acute{\text{e}}\)clet number~$Pe_\psi$.
Increasing $Pe_\psi$ from 1 to 10 strengthens the suppression, whereas a further increase to 100 weakens it.
Since Marangoni stresses arise from interfacial surfactant concentration gradients, this trend shows that $Pe_\psi$ influences coarsening indirectly by controlling the buildup and persistence of those gradients, rather than directly setting the stress magnitude.


To determine whether this non-monotonicity originates from bulk transport, we examine the average convective and diffusive fluxes of the phase field, $J_{\phi,c} = \langle|\phi \bf{u}|\rangle$ and $J_{\phi,d}=\frac{1}{Pe_\phi}\langle|{\nabla}\mu_\phi|\rangle$~\citep{tanaka1998spontaneous}.
As shown in Fig.~\ref{fgr:5}(b), coarsening is dominated by hydrodynamic convection, with $J_{\phi,c}$ (solid lines) consistently exceeding $J_{\phi,d}$ (dashed lines) throughout the coarsening stage.
More importantly, both $J_{\phi,c}$ and $J_{\phi,d}$ nearly collapse for different $Pe_\psi$, indicating that the bulk transport of the phase field remains essentially unchanged.
This demonstrates that the non-monotonic evolution of $R(t)$ cannot be attributed to the bulk coarsening dynamics of $\phi$, and instead points to an interfacial origin associated with surfactant transport.

To identify this interfacial origin, we next examine the temporal evolution of the average capillary and Marangoni forces in Figure~\ref{fgr:5}(c).
The capillary force $\langle|\mathbf{F_{Ca}}|\rangle = \langle|\frac{1}{Ca}\frac{3C n}{\sqrt{8}} ( \nabla \cdot \overline{\boldsymbol{\tau}}+ \beta_s \ln (1-\psi) \nabla \cdot \overline{\boldsymbol{\tau}})|\rangle$ (solid lines) exhibits a sharp pulse around $t \approx 6$, marking the onset of rapid interface coalescence and domain growth.
This capillary pulse is nearly identical for all three values of $Pe_\psi$, indicating that the the early-stage capillary-driven dynamics during are insensitive to surfactant transport.
By contrast, the Marangoni force $\langle|\mathbf{F_{Ma}}|\rangle = \langle|Ma\frac{3Cn}{\sqrt{8}} \nabla \ln (1-\psi) \cdot \overline{\boldsymbol{\tau}}|\rangle$ (dashed lines) exhibits  a delayed and strongly $Pe_\psi$--dependent response.
For $Pe_\psi=1$, the Marangoni response follows the capillary pulse with a slight delay and develops a secondary peak during the onset of coarsening suppression (gray shaded region). 
For $Pe_\psi=10$, this secondary peak becomes both stronger and more sustained, leading to the strongest suppression of domain growth. 
For $Pe_\psi=100$, although advection continues to generate interfacial non-uniformity, the Marangoni response weakens because adsorption-desorption can no longer replenish the interface rapidly enough, resulting in reduced interfacial surfactant availability. 
These results show that the suppression of coarsening is governed not simply by the presence of Marangoni stresses, but by how their spatiotemporal evolution is selected by surfactant transport.

\begin{figure}[h!]
  \centering
  \includegraphics[width=1.0\textwidth]{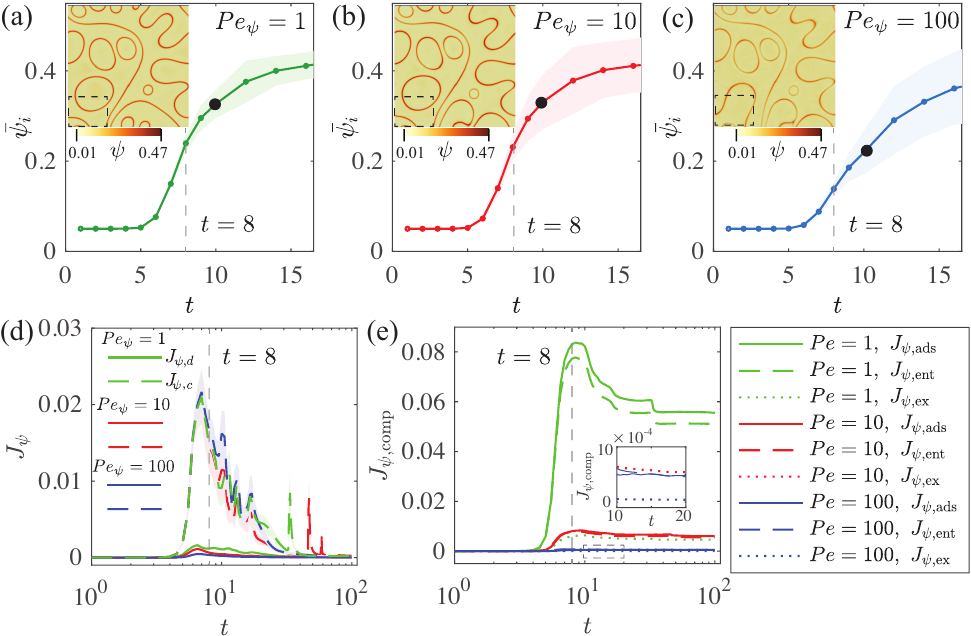}
  \caption{
    Surfactant mass transport for different surfactant P\(\acute{\text{e}}\)clet numbers $Pe_\psi$. 
    (a--c) Mean interfacial surfactant concentration $\bar{\psi_{\mathrm{i}}}$ for $Pe_\psi=1$, $Pe_\psi=10$, and  $Pe_\psi=100$. 
    The light-shaded band represents standard deviation of $\bar{\psi_{\mathrm{i}}}$, showing the degree of spatial non-uniformity of interfacial surfactant concentration $\psi_{\mathrm{i}}$.
    The dashed lines indicate the time instant~$t = 8$ and the inset shows the surfactant concentration field at $t = 10$.
    (d) Average convective flux $J_{\psi,c}$ and the average diffusive flux $J_{\psi,d}$ of surfactants as functions of time.
    The light-shaded band represents standard deviation of $J_{\psi,c}$ and $J_{\psi,d}$, and these values are multiplied by 0.15 for clarity.
    (e) Time evolution of the components of the average diffusive surfactant flux~$J_{\psi,d}$, i.e.,~the adsorption flux~$J_{\psi,ads}$,  entropy flux~$J_{\psi,ent}$, and enthalpy flux $J_{\psi,ex}$.
    }
  \label{fgr:6}
\end{figure}

To further clarify this transport origin, we examine the temporal evolution of the mean interfacial surfactant concentration, $\bar{\psi}_{\mathrm{i}}$, its standard deviation, and the associated surfactant fluxes shown in Fig.~\ref{fgr:6}. 
For $Pe_\psi=1$ and 10, the mean interfacial concentration follows a similar evolution and reaches a comparable plateau value, whereas the standard deviation is noticeably larger for $Pe_\psi=10$, indicating stronger interfacial heterogeneity. 
For $Pe_\psi=100$, by contrast, the interfacial distribution remains highly non-uniform, but the mean interfacial concentration is substantially reduced.
The interfacial concentration fields at $t=10$ (insets of Figs.~\ref{fgr:6}(a-c)) confirm this distinction.
The $Pe_\psi=10$ case maintains both strong heterogeneity and substantial interfacial loading, whereas the $Pe_\psi=100$ case exhibits strong heterogeneity on an overall depleted interface.

%
%

To further validate this interpretation, we examine the temporal evolution of the mean interfacial surfactant concentration, $\bar{\psi}_{\mathrm{i}}$, its standard deviation, and the associated surfactant fluxes shown in Fig.~\ref{fgr:6}. 
As shown in Figs.~\ref{fgr:6}(a--c), the mean interfacial concentration increases rapidly after $t\approx 6$ in all cases. 
For $Pe_\psi=1$ and $Pe_\psi=10$, $\bar{\psi}_{\mathrm{i}}$ follows a similar temporal evolution and reaches a comparable plateau value. 
However, the standard deviation is noticeably larger for $Pe_\psi=10$, indicating a stronger spatial non-uniformity of surfactant concentration along the interface. 
By contrast, for $Pe_\psi=100$, although the standard deviation remains large and the interfacial distribution is highly heterogeneous, the mean interfacial concentration stays substantially lower than in the other two cases. 
The interfacial concentration fields at $t=10$ (insets) further support this trend. 
Compared with $Pe_\psi=1$, the $Pe_\psi=10$ case preserves stronger interfacial heterogeneity while maintaining a relatively high surfactant level, whereas the $Pe_\psi=100$ case exhibits pronounced spatial variation but an overall depleted interface.

The flux budget shown in Figs.~\ref{fgr:6}(d,e) further clarifies the origin of this behavior. 
Figure~\ref{fgr:6}(d) shows that the convective flux, $J_{\psi,c}=\langle |\psi \mathbf{u}| \rangle$, 
exhibits a sharp peak around $t\approx 7$--$8$ and is substantially larger than the diffusive flux, $J_{\psi,d}=\frac{1}{Pe_\psi}\left\langle \left| M(\psi)\nabla\mu_\psi \right| \right\rangle$, for all three cases. 
This indicates that, during the transient stage, advection mainly redistributes surfactant along the interface. 
By contrast, the diffusive transport governs the replenishment of surfactant from the bulk to the interface and depends strongly on $Pe_\psi$.

To elucidate this diffusive contribution, we decompose the surfactant chemical potential in Eq.~\eqref{555} as
$\mu_\psi=\mu_{\psi,\mathrm{ent}}+\mu_{\psi,\mathrm{ads}}+\mu_{\psi,\mathrm{ex}}$.
Accordingly, the diffusive surfactant flux can be decomposed into three contributions~\citep{SOLIGO20191292,yang2018numerical},
\begin{equation*}
  \label{eq:energy_budget2}
   \underbrace{\frac{1}{Pe_\psi}\left\langle \left| M(\psi)\nabla\mu_\psi \right| \right\rangle}_{J_{\psi,d}} = \underbrace{\frac{1}{Pe_\psi}\left\langle \left| M(\psi)\nabla \mu_{\psi,\mathrm{ent}} \right| \right\rangle}_{J_{\psi,\mathrm{ent}}} + \underbrace{\frac{1}{Pe_\psi}\left\langle \left| M(\psi)\nabla \mu_{\psi,\mathrm{ads}} \right| \right\rangle}_{J_{\psi,\mathrm{ads}}}+ \underbrace{\frac{1}{Pe_\psi}\left\langle \left| M(\psi)\nabla \mu_{\psi,\mathrm{ex}} \right| \right\rangle}_{J_{\psi,\mathrm{ex}}},
\end{equation*}
with $\mu_{\psi,\mathrm{ent}} = Pi\ln\left(\frac{\psi}{1-\psi}\right)$,
$\mu_{\psi,\mathrm{ads}} = -S_{ad}(1-\phi^2)^2$,
$\mu_{\psi,\mathrm{ex}} = E_x\phi^2$.
Here, $J_{\psi,\mathrm{ent}}$ is the entropy-driven diffusive contribution, which tends to homogenize the surfactant concentration and smooth interfacial concentration gradients, $J_{\psi,\mathrm{ads}}$ is the adsorption-driven diffusive contribution, which transports surfactant toward the diffuse interface and replenishes interfacial coverage, and $J_{\psi,\mathrm{ex}}$ is the enthalpy-driven diffusive contribution associated with the energetic penalty of free surfactant in the bulk phase. 
Since the absolute value is used in the flux definitions, these quantities measure the magnitude of each transport contribution rather than the signed net flux.

As shown in Fig.~\ref{fgr:6}(e), the diffusive transport is dominated by the adsorption (solid lines) and entropy (dashed lines) contributions, whereas the enthalpy contribution (dotted lines) remains negligible throughout. 
This hierarchy reflects the fact that surfactant transport is controlled primarily by interfacial adsorption and gradient relaxation, while the bulk enthalpic penalty plays only a minor role under the present conditions. 
For $Pe_\psi=1$ (green lines), both adsorption- and entropy-driven contributions are strong: the former efficiently replenishes surfactant at the interface, whereas the latter strongly smooths interfacial concentration gradients. 
For $Pe_\psi=10$ (red lines), adsorption remains sufficiently effective to maintain interfacial loading, while entropy-driven smoothing is weaker, allowing stronger spatial non-uniformity to persist and thereby sustaining a larger and more persistent surface-tension gradient. 
For $Pe_\psi=100$ (blue lines), both adsorption- and entropy-driven diffusive contributions become much weaker. 
As a result, although advection still generates pronounced interfacial heterogeneity, surfactant replenishment from the bulk is insufficient, leading to an overall depleted interface and hence a weaker Marangoni response.

These results demonstrate that the $Pe_\psi=10$ case achieves the most favorable balance between interfacial surfactant replenishment and gradient retention. In this regime, sufficient interfacial loading coexists with persistent concentration gradients, producing the strongest Marangoni response and, consequently, the strongest suppression of hydrodynamic coarsening.

\section{Conclusions}
\label{sec6}

We have investigated how soluble surfactants regulate bicontinuous liquid-liquid phase separation in the hydrodynamic regime using a two-order-parameter phase-field model coupled to the incompressible Navier--Stokes equations. 
The numerical framework was first validated against the classical diffusive and hydrodynamic coarsening scalings and against benchmark results for surfactant-laden droplet deformation, establishing its accuracy for the present problem.

The simulations show that soluble surfactants suppress hydrodynamic coarsening primarily through Marangoni stresses generated by interfacial surfactant concentration gradients. 
By comparing clean systems, surfactant-laden systems with Marangoni stresses, and surfactant-laden systems with Marangoni stresses artificially removed, we demonstrate that the dominant suppression mechanism is not the reduction of mean interfacial tension, but the Marangoni-driven redistribution of interfacial momentum. 
These stresses hinder the drainage of thin liquid films, delay coalescence between approaching interfaces, and alter the local flow organization, thereby changing both the coarsening rate and the morphological evolution pathway of bicontinuous domains.

A central result of this work is that the Marangoni-mediated suppression of coarsening exhibits a non-monotonic dependence on the surfactant P\'eclet number, $Pe_\psi$. 
This behavior reflects a competition between interfacial replenishment and gradient retention.
Sustained Marangoni stresses require both sufficient surfactant supply, maintained by diffusive exchange between the bulk and the interface, and persistent interfacial concentration gradients, which must survive diffusive smoothing.
At low $Pe_\psi$, diffusion readily replenishes the interface  suppresses gradient formation.
At high $Pe_\psi$, advection preserves interfacial heterogeneity, yet interfacial surfactant supply becomes insufficient. 
The maximum suppression therefore occurs at an intermediate $Pe_\psi$, where interfacial coverage and gradient persistence are most favorably balanced.

These findings provide a mechanistic framework for understanding transport-controlled Marangoni regulation of bicontinuous hydrodynamic coarsening. 
More broadly, they show that soluble surfactants do not simply reduce interfacial tension passively, but dynamically select coarsening pathways through their coupled interfacial transport and hydrodynamic feedback. 
This insight may prove useful for tuning microstructure evolution in emulsions, polymer blends, and other surfactant-laden multiphase systems~\citep{Cates_Tjhung_2018, Wang2019,LI2023299}. 
Future work could extend the present framework to include interfacial rheology~\citep{manikantan2020surfactant, Ewoldt2022}, asymmetric material properties~\citep{hardy2026kinetic}, and multicomponent mixtures~\citep{Li_Wan_Hao_Diddens_Zhang_Tan_2025,zwicker2025physics}.

\appendix

\section{Dimensionless free energy functional}
\label{sec:sample:appendixa}

The dimensionless free-energy functional consists of four contributions, given by
\begin{equation}\label{111}
  \begin{aligned}
    F(\phi, \psi)= & \int_{\Omega}[\underbrace{ {\frac{1}{4} (\phi^2-1)^2+\frac{Cn^2}{2}|\nabla \phi|^2}}_{f_\phi}+\underbrace{Pi[\psi \ln \psi+(1-\psi) \ln(1-\psi)]}_{f_\psi}\\
    &-\underbrace{S_{ad} \psi (1-\phi^2)^2}_{f_a}+\underbrace{E_x \psi \phi^2}_{f_{ex}} ]~d \Omega,
    \end{aligned}
\end{equation}
where $\Omega$ is the computational domain and $Cn$ is the Cahn number. 
Here, the first contribution, $f_\phi$, is the Ginzburg-Landau free-energy density for binary mixtures.
It includes a classical double-well potential, which determines the bulk free-energy landscape, together with a gradient term that accounts for interfacial energy~\citep{Engblom2013,LIU20109166,SOLIGO20191292}. 
The two minima of double-well potential term at $\phi=\pm 1$ correspond to the equilibrium states of the two pure immiscible components (e.g., water and oil)~\citep{Cates_Tjhung_2018}.
The second contribution, $f_\psi$, follows the Flory-Huggins form and represents the entropic free energy associated with mixing surfactant molecules with the bulk phase.
Surfactants preferentially adsorb at interfaces, reducing interfacial free energy.
This effect is captured by the adsorption term $f_a$ which introduces a negative energetic contribution that drives surfactant accumulation at the interface.
The last term, $f_{ex}$, accounts for the enthalpic contribution of free surfactant in the bulk phase.
In addition, the parameters $Pi$, $S_{ad}$, and $E_x$ correspond, respectively, to the temperature-dependent coefficient of the entropy term~$f_\psi$, the coefficient of the adsorption term~$f_a$, and the coefficient of the enthalpy term~$f_{ex}$.

\section{Choice of parameters}
\label{sec:sample:appendixb}
The dimensional parameter ranges of the emulsion system are given in Table~\ref{tbl:example1}. 
Here, the characteristic length and velocity $L$ are taken as  $L \sim 10^{-5} - 10^{-4}~\mathrm{m}$ and $U \sim 10^{-2}- 10^{-1}~\text{m/s}$~\cite{hester2023}, respectively.
The reference values of surface tension $\sigma_0$, density $\rho$, and dynamic viscosity $\eta$ are chosen based on typical water-mineral-oil systems.
Using the dimensional ranges in Table~\ref{tbl:example1}, the Reynolds number is estimated as $Re = \rho UL / \eta \sim 10^{-2} - 10$ and the capillary number as $Ca = \eta U / \sigma_0 \sim 10^{-3} - 0.1$.
In Sections~\ref{sec4} and \ref{sec5}, we adopt representative values of $Re = 0.1$ and $Ca = 0.1$, both of which fall well within the physically relevant range for LLPS emulsions.


\begin{table}[h]
\centering
  \small
    \caption{Range of values of physical parameters}
    \label{tbl:example1}
    \begin{tabular*}{0.9\textwidth}{@{\extracolsep{\fill}}cccc}
      \hline
      Dimensional parameters & Symbol & Value range & Units\\
      \hline
      Characteristic length & $L$ & $10^{-5} \sim 10^{-4}$ & $\mathrm{m}$\\
      Characteristic velocity & $U$ & $10^{-2} \sim 10^{-1}$ & $\mathrm{m\cdot s^{-1}}$\\ 
      Surface tension & $\sigma_0$ & $10^{-2}$ & $\mathrm{kg \cdot s^{-2}}$ \\
      Density  & $\rho$ & $10^{3}$ & $\mathrm{kg \cdot m^{-3}}$\\
      Dynamic viscosity & $\eta$ &  $10^{-3} \sim 10^{-2}$ & $\mathrm{kg\cdot m^{-1}\cdot s^{-1}}$\\               
      Diffusion coefficient($\phi$) & $D_{\phi_{,} \text {Fick }}$ &  $10^{-9}$ & $\mathrm{m^2 \cdot s^{-1}}$\\
      Diffusion coefficient($\psi$)~\cite{chang1995adsorption,weinheimer1981diffusion,ribeiro2003diffusion} & $D_{\psi_{,} \text {Fick }}$ &  $10^{-11} \sim 10^{-8}$ & $\mathrm{m^2 \cdot s^{-1}}$\\
      \hline
    \end{tabular*}
\end{table}


The $Pe_\phi$ and $Pe_\psi$ in Eqs.\eqref{222} and \eqref{333} are given by~\cite{LIU20109166}
\begin{equation}\label{16}
  \begin{aligned}
 & Pe_\phi=Pe^*_\phi \frac{\partial \mu_{\phi}}{\partial\phi} \approx Pe^*_\phi (3 \phi^2-1), \\
   \end{aligned}
\end{equation}
\begin{equation}\label{17}
  \begin{aligned}
 & Pe_\psi=Pe^*_\psi M(\psi) \frac{\partial \mu_{\psi}}{\partial\psi} = Pe^*_\psi Pi, \\
   \end{aligned}
\end{equation}
where $Pe^*_\phi=\frac{U L }{D_{\phi_{,} \text {Fick }}} \sim 10^2 - 10^4$ and $Pe^*_\psi=\frac{U L }{D_{\psi_{,} \text {Fick }}} \sim 1 - 10^5$. 
In this work, we fix $Pe_\phi=100$ (correspond to $Pe^*_\phi \sim 300$) in Eq.\eqref{222}. 
For the surfactant field $\psi$, we consider $Pe_\psi \sim 1 - 10^2$ (corresponding to $Pe^*_\psi \sim 10 - 10^3$) to explore the influence of surfactant diffusivity.

To accurately resolve the surfactant concentration and interfacial tension profiles across the diffuse interface, we follow Ref.~\citep{LIU20109166} and set the Cahn number to $Cn=0.04 \approx 2 d x$,~which corresponds to an effective interfacial thickness~$\delta \approx 0.17$.
With this choice, the interfacial P\(\acute{\text{e}}\)clet number $Pe_\phi\delta < 590$, below the critical threshold for diffuse-interface Marangoni instability~\citep{Li_Wan_Hao_Diddens_Zhang_Tan_2025}.
It should be noted that the choice of $Cn$ also influences the diffusive dynamics of phase separation modifying the width of the interfacial region, since the negative diffusion coefficient of $\phi$ arises only within this region (Eq.~\eqref{16}). 
However, this does not alter the coarsening rate of Model~B~\citep{Bray01061994}.
Moreover, when the time scale of diffusion~($t_d \sim {L}^2/D_\phi$) exceeds the hydrodynamic time scale~($t_h \sim L/U$), the influence of $Cn$ on the scaling behavior of Model~H is also negligible~\cite{hester2023}.

 \bibliographystyle{elsarticle-num} 
 \bibliography{cas-refs}





\end{document}